# Thermal Simulation and Experimental Analysis of Optically Pumped InP-on-Si Micro- and Nanocavity Lasers

Pengyan Wen*, Preksha Tiwari, Markus Scherrer, Emanuel Lörtscher, Bernd Gotsmann and Kirsten E. Moselund

IBM Research Europe – Zurich, Saeumerstrasse 4, CH-8803 Rueschlikon, Switzerland

* Email: pew@zurich.ibm.com

**ABSTRACT:** There is a general trend of downscaling laser cavities, but with high integration and energy densities of nanocavity lasers, significant thermal issues affect their operation. The complexity of geometrical parameters and the various materials involved hinder the extraction of clear design guidelines and operation strategies. Here, we present a systematic thermal analysis of InP-on-Si micro- and nanocavity lasers based on steady-state and transient thermal simulations and experimental analysis. In particular, we investigated the use of metal cavities for improving the thermal properties of InP-on-Si micro- and nanocavity lasers. Heating of lasers is studied by using Raman thermometry and the results agree well with simulation results, both reveal a temperature reduction of hundreds of Kelvins for the metal-clad cavity. Transient simulations are carried out to improve our understanding of the dynamic temperature variation under pulsed and continuous-wave pumping conditions. The results show that the presence of a metal cladding not only increases the overall efficiency in heat dissipation, but also causes a much faster temperature response. Together with optical experimental results under pulsed pumping, we conclude that a pulse width of 10 ns and repetition rate of 100 kHz is the optimal pumping condition for a 2 μm wide square cavity.

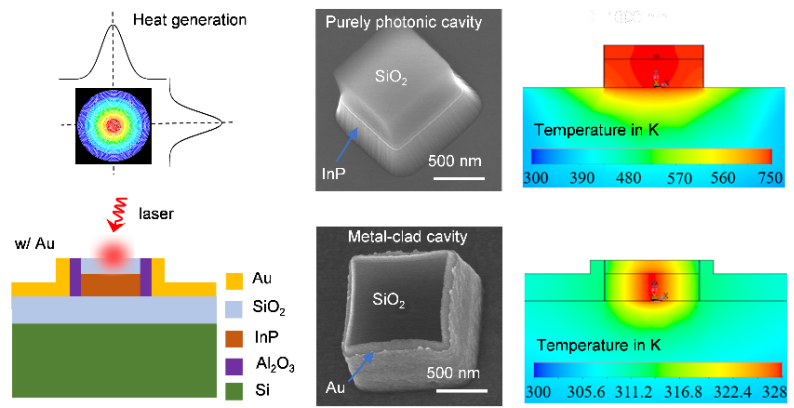

Keywords: thermal simulation; heating effects; Raman spectroscopy; optically pumped nanolaser; InP-on-Si; metal-clad cavity.

## INTRODUCTION

Semiconductor micro- and nanocavity lasers have attracted much attention due to their high-quality factor (Q-factor), low-threshold lasing and low power consumption, promoting them as excellent candidates for light sources in optical interconnects for large scale integration.[1-3] In particular, InP-on-Si nanocavity lasers with low-loss whispering gallery modes have been widely used in dense photonic integrated circuits (PICs), optical communication and optical sensing.[4-6] InP is a common material for integrated photonics because it enables the growth of lattice-matched InGaAs quantum wells (QWs) for high performance lasers at telecom wavelengths. Multiple aspects of nanocavity laser design including geometry, electromagnetic properties, and surface passivation have been investigated in order to achieve resonances with high Q-factor and low-loss modes, realizing nanocavity lasers with low power consumption and long-term stability.[7-10] Whereas shrinking the cavity size can improve efficiency of nanocavity lasers compared to their larger counterparts, it can also lead to excessive heating due to the high power density. Thermal effects are known to be key factors affecting the performance and long-term stability, even causing degradation and failure of nanocavity lasers. Yet, thermal studies have received much less attention than electromagnetic or other optical ones.

Metal-clad nanocavities were first designed in 2007 to achieve device sizes smaller than the free-space wavelength due to their tight confinement of optical modes.[11,12] Improved size scaling potential was demonstrated for such metal-clad lasers, but the metal also introduced absorption loss which tends to increase lasing threshold and be a cause of non-radiative loss. Since then, there is also some debate whether such modes are plasmonic, merely photonic or so-called hybrid photonic-plasmonic in nature as they benefit from increased reflection and confinement provided by the metal. Various types of metal-clad microcavity lasers have been demonstrated ranging from pulsed optically pumped or continuous wave (cw) electrically pumped.[13-15] Although it is expected that a metallic cavity will improve heat dissipation in nanocavity lasers, so far there has been no systematic study on the thermal analysis of dielectric and metallic micro- and nanocavity lasers. Effects of the substrate thickness, heat spreading and pump conditions on the thermal properties of laser were simulated and results showed that a thin substrate and a heat spreader significantly enhance heat dissipation.[16] More detailed heat dissipation in electrically pumped metallic-dielectric nanolasers with $Al_2O_3$ and $SiO_2$ dielectric shield were compared, showing better performance for the $Al_2O_3$ dielectric shield with high thermal conductivity.[17] More recently, thermal effects of InP/InAs nanowire lasers integrated on different optical platforms were studied and results revealed that decreasing the thermal resistance between the nanowires and substrates is the most efficient way for cw nanowire laser operation.[18] There are also few studies focusing on the transient thermal effects in semiconductor lasers.[19, 20] Size scalability of micro- and nanocavity lasers is another important consideration as wavelength- and sub-wavelength scale lasers have been pursued to realize smaller footprints, lower lasing thresholds and higher modulation speeds, giving rise to non-classical phenomena which are hardly seen in conventional resonators, such as the Purcell effect (enhancement of the spontaneous emission rate) and β-factor (spontaneous emission coupling to the lasing mode).[21-23] Furthermore, size scaling leads to an increased effect of thermal interfaces that need to be taken into account.

Even though the ability to operate under cw conditions would be desirable for a stable and continuous power output, in reality most nanocavity lasers demonstrated are operating under pulsed excitation.[24] Therefore, another important consideration for the thermal behavior of nanocavity lasers are the transient thermal effects, as the time for the temperature to reach a steady state is often long compared to the operating frequency.[25] The transient pumping power to reach a reasonable average lasing power, is usually several times higher under pulsed than under cw operation, thus leading to high transient temperature peaks. To evaluate the dynamic laser performance and establish ideal pumping conditions with optimal pulse width and frequency, dynamic thermal effects need to be studied.

In this paper, we systematically study steady-state and dynamic thermal effects of optically pumped InP-on-Si micro- and nanocavity lasers with and without a metal cladding by both simulation and experimental validation. We evaluate the impact of a metal cladding, cavity-size as well as the dielectric environment on the thermal budget. Temperature distributions of purely photonic (without metal) and metal-clad plasmonically-enhanced micro- and nanocavity (with Au) lasers are studied by both steady-state simulation and experimental Raman characterization. The influence of cavity shape, cavity size and dielectric layer thickness are investigated, and guidelines for cavity design are provided from a thermal perspective. Dynamic thermal effects with and without Au cladding are also investigated for various pulse widths and frequencies. The transient thermal simulation results together with the experimental lasing thresholds of a purely dielectric square cavity of 2 μm diameter, both reveal that a pulsed pumping with 10 ns pulse width and 100 kHz frequency is optimal. Furthermore, the Au-clad cavity shows a much faster temperature response.

RESULTS
A. Steady-state thermal analysis

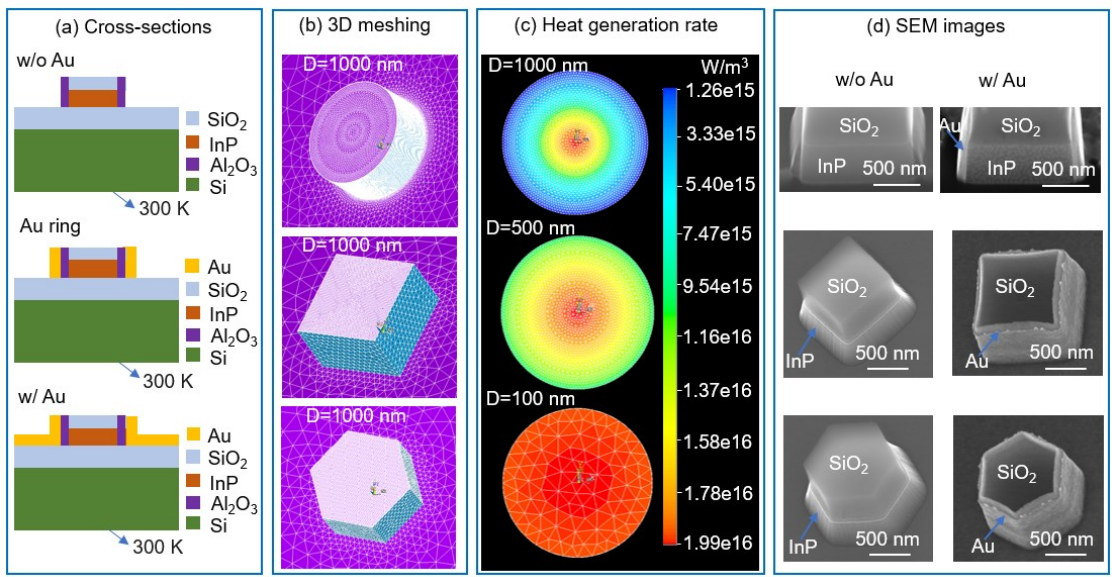

Figure 1. (a) Schematic cross-sections of nanocavity lasers w/o Au, with isolated Au ring and w/ Au (Au cladding covering the surface), an initial uniform temperature of 300 K and backside temperature of the substrate of 300 K are used in the simulation. (b) Meshing model of the disk, square and hexagon cavities with a diameter of 1000 nm. (c) Overlay of the heat generation rate (in $W/m^3$) with meshing structure

in micro- and nanodisk lasers with diameter of 100 nm, 500 nm and 1000 nm in the same scale bar used for both steady-state and transient simulations. (d) Cross-section (top row) and tilted top-view (middle and lower rows) scanning electron microscope (SEM) images of square and hexagonal cavities w/o and w/ Au cladding.

**Steady-state simulations.** The steady-state simulations are carried out using ANSYS Parametric Design Language (APDL). Various geometries including disk, square and hexagon are studied, with diameters ranging from 100 nm to 2 μm. We chose these different geometries on the following grounds: the circular disk is the most common whispering-gallery mode cavity, hexagons form naturally when we grow our III-V micro- and nanocavity lasers monolithically on Si (111),[24] and squares are the simplest geometry from a simulation perspective. The schematic cross-sections, meshing models, heat generation rate and scanning electron microscope (SEM) images are shown in Fig. 1 (a), (b), (c) and (d), respectively. The Si substrate considered in the simulation environment is 50 μm × 50 μm × 50 μm with a 2 μm-thick $SiO_2$ layer on top of it. The in-plane thickness of Au is 250 nm, while on the sidewalls of the disk it is 150 nm. The thickness of the InP disk is 300 nm with 150 nm $SiO_2$ (forming from exposed and developed photoresist - HSQ) covering the top facet.

Between the disk and the Au there is a 5 nm $Al_2O_3$ layer intended to reduce absorption losses from the metal. These base parameters for simulation are chosen to reflect the actual specifications of the fabricated devices, except for the Si substrate which is chosen sufficiently large to not be a limiting factor in the simulation. An overview of the 3D model used in the simulation is shown in Supporting Information 1. Fig. 1(d) shows the SEM images of square and hexagon cavities which we measured experimentally.

In the simulations, the lasers are optically pumped with a spatial-dependent Gaussian distributed heat generation load with a total integrated power of 1.69 mW and a diameter of 1 μm, hitting the center of the cavities. This power corresponds to the experimentally found situation slightly below the lasing threshold for the majority of our micro- and nanocavity lasers, so we assume the total optical pump power is converted into heat in this regime of operation. More details on the experimental conditions and consideration of the heat convection influence on the temperature are discussed in the Supporting Information 1. Thermal parameters of the materials used in this paper are listed in Table 1.[26-29] These are typical parameters used in bulk materials. Details on the selection of these parameters and the heat transfer models are discussed in the Methods section. In the simulation, we take the thermal boundary conductance at each interface into account, with the values listed in Table 2.[30-34]

Table 1. Thermal parameters used in the model [26-29]

| Material | Thermal conductivity (W/m·K) | Specific heat (J/kg·K) | Density (kg/m³) |
|---|---|---|---|
| Si | 131 | 700 | 2320 |
| $SiO_2$ | 1.4 | 440 | 2200 |
| InP | 68 | 180 | 4810 |
| $Al_2O_3$ | 17 | 880 | 3890 |
| Au | 318 | 128 | 19320 |

Table 2. Thermal boundary conductance used in the model [30-34]

| Interfaces | Thermal boundary conductance (W/m²·K) |
|---|---|
| $SiO_2$ / $Al_2O_3$ | $4.0 \times 10^8$ |
| InP / $Al_2O_3$ | $2.2 \times 10^8$ |
| $SiO_2$ / Au | $5.3 \times 10^7$ |
| $Al_2O_3$ / Au | $1.2 \times 10^8$ |

We first studied the role of the Au cladding on the heat dissipation by comparing the temperatures of three typical cavities shown in Fig. 1(a). Fig. 2(a) shows the highest temperature at the center of the cavities with various diameters, all show a significant reduction of temperature by using Au cladding. Even using only an Au ring around the perimeter of the cavity sidewalls could reduce the highest temperature by approximately half. Fig. 2(b) shows the thermal conductance of the disk lasers with various diameters. The thermal conductance of the cavity with Au is much higher than the one without Au, however, the difference starts to level out as the diameter increases to 1000 nm. In addition, the Au plays a significant role for efficient heat dissipation in devices with larger diameters. Fig. 2(c), (d) and (e) show the temperature distribution in the cross-section of the nanodisk laser. For the cavity with Au covering the entire substrate surface, the Au acts like a heat sink. In the case with only an Au ring surrounding the cavity, it still improves heat removal significantly by transferring heat to the underlying layers. In the cavity without Au, the heat can only be transferred down to the $SiO_2$ layer through the area determined by the diameter of the disk. Due to the small thermal conductivity of $SiO_2$, heat cannot be transferred efficiently to the substrate, thus resulting in hundreds of Kelvins higher temperature in the InP layer. More detailed simulation results as a function of the Au dimension can be found in the Supporting Information 2.

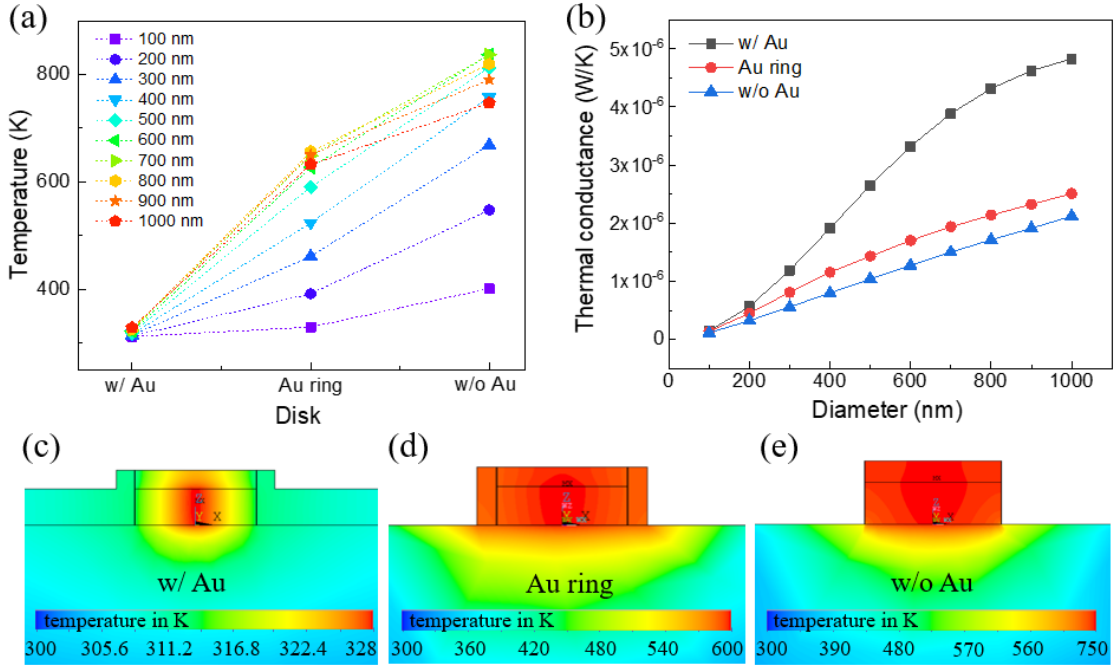

Figure 2. Steady-state simulation results on micro- and nanodisk cavities. (a) Highest temperature at the center of the cavities with various diameters, pumped with a Gaussian distributed optical power with a total power of 1.69 mW. (b) Thermal conductance of the disk laser with various InP diameters from 100 nm to 1000 nm, defined as the total heat flux $\dot{Q}$ divided by the highest $\Delta T$. (c) Temperature distribution for disk laser with Au cladding. (d) Temperature distribution for disk laser with Au ring. (e) Temperature distribution for disk laser w/o Au (purely photonic). Note the significant difference in the scale in the three designs.

In addition to the metal-clad design, we also studied the influence of the InP cavity shapes and thickness of the SiO$_2$ on the thermal properties. Fig. 3(a) shows the highest temperature of the disk, square, and hexagon cavities with various diameters, both with and without Au. We consider only the case where Au is also covering the substrate around the cavity in this figure as this corresponds to our experimental devices. We see a large temperature difference of hundreds of Kelvins between the Au-clad cavity and the purely photonic cavity while a relatively minor temperature difference for different shapes. Moreover, the disk and hexagon lasers without Au shows a maximum temperature at diameter around 700 nm which can be ascribed to the total heat saturation under the Gaussian distribution. The maximal temperature for the square cavities is found around 800 nm side length. To better understand the results in Fig. 3(a), the heat flux in both cavities is presented in Figs. 3 (c) and (d), which shows a heat flux value being five times larger for the Au-clad device.

We also compared the temperature distribution in different parts of the cavities and found the highest temperature gradient is located in the bottom SiO$_2$ layer, see details in Supporting Information 1. So here we compare temperatures of cavities with various thickness of the bottom SiO$_2$ layer shown in Fig. 3 (b). While for cavities w/ Au cladding, the maximum temperatures drastically decrease with decreasing SiO$_2$ thickness, the cavities w/o Au cladding reveal a difference that is less. This is because the Au cladding helps the removal of heat by draining the heat to a larger parallel area. Especially for lasers with large diameters, a thinning of the SiO$_2$ box provides an efficient mean for reducing the temperature increase while maintaining an optimum dielectric isolation for the optical mode. Based on our experiments on template-assisted selective epitaxy (TASE) grown InP on Si, it should be possible to reduce this to a thickness as small as 300nm without significantly degrading the Q-factor of the cavity.[24]

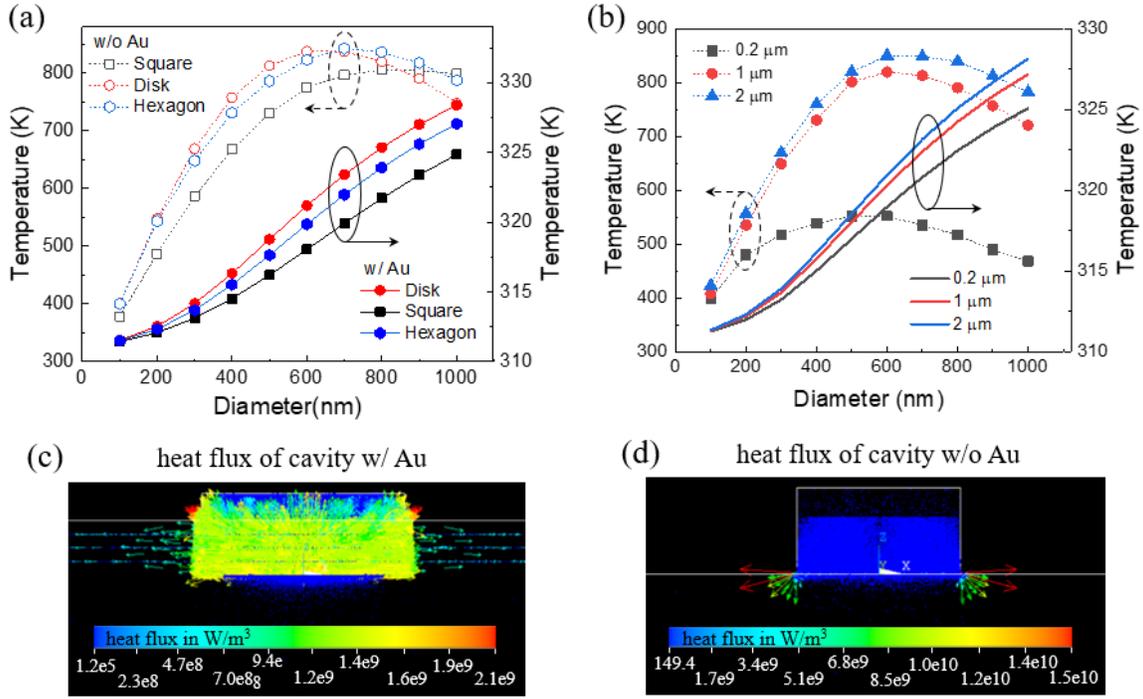

Figure 3. Steady-state simulation results with total heat flux $\dot{Q}$ of 1.69 mW. (a) Temperature dependence on diameter for nanocavity lasers with various shapes, the purely photonic cavities show highest temperature at diameter of around 700 nm while the temperature continuously increase as the diameter increase in the Au-clad cavities. (b) Temperature dependence on diameter for various bottom SiO$_2$ thickness for w/o Au and w/ Au cavities. (c) Cross-sectional heat flux image for a 1 μm diameter w/ Au cavity. (d) Cross-sectional heat flux image for a 1 μm diameter w/o Au cavity.

The thickness of the Al$_2$O$_3$ is also studied and the results are shown in the Supporting Information 3. The highest temperatures of the 2 nm thick Al$_2$O$_3$ is slightly higher than that of 5 nm thickness at various diameters for cavities both with and without Au, indicating only a minor significance of the Al$_2$O$_3$ layer thickness for the thermal design. However, for the optical mode this thickness is quite important. With a direct Au-InP interface, a pure plasmonic mode might be possible, but generally it is found that by using a thin interlayer dielectric the majority of the mode energy will be confined in a hybrid photonic-plasmonic mode, which is found to significantly reduce the optical loss introduced by the metal.[35, 36] The temperature-dependent thermal conductivity is considered, and the results are compared with the constant thermal conductivity in the Supporting Information 4.

**Temperature characterization using Raman thermometry.** To verify the steady-state thermal simulation results, we characterized the temperature of micro- and nanocavities by experimentally conducting Raman thermometry studies. Such measurements are made possible by the fact that the Raman set-up used is equipped with a cooling/heating stage that covers the temperature range from 78 K to 900 K. This enables a calibration of the specific structure's peak broadening over a wide temperature range. The fact that the system is located in an ultra-silent lab[37] makes long integration times on the order of hundreds of seconds with negligible spatial shifts of the sample and thermal drifts of the monochromator and other optical components possible. The InP nanocavities studied here are fabricated by wafer bonding of 300 nm InP on SiO$_2$ on silicon, subsequent etching using a HSQ mask and inductively coupled plasma (ICP) dry etching, followed by a short wet etch using a 1:10 diluted phosphoric acid solution to clean the surface. After E-beam exposure and development, HSQ turns into a SiO$_2$-like oxide. We keep this oxide on top of the nanocavities to provide better dielectric-constant matching to the underlying SiO$_2$. The oxide is transparent to the wavelength used in our experiment. The devices are coated by 3 nm Al$_2$O$_3$ oxide layer followed by electron-beam evaporation of 250 nm Au (first 150 nm evaporated under a tilt angle of 45 degrees, remaining 100 nm evaporated without tilting of the sample stage) using a 4 nm thin Ti adhesion layer in between. The Au is subsequently removed from the top facet by Ar-ion milling after protecting the side walls with unexposed HSQ which is removed after the milling process.

The local temperature of InP-on-Si nanolasers are measured by Raman thermometry in back-scatter reflectance geometry and a spot size below 1 μm. As a steady-state technique for temperature measurements, there are three temperature-dependent optical sensing modalities which can be used for probing the average temperature within the probing volume (penetration depth at 532 nm larger than the InP layer thickness): Raman Stokes' peak shift, ratio of anti-Stokes and Stokes intensities, and Raman Stokes peak width broadening (full width at half maximum, FWHM) of a specific phonon mode. The Stokes' peak shift is the simplest and easiest way to determine the temperature, a temperature increase causes thermal expansion and lowers the interatomic forces within the lattice, which is reflected in a shift towards lower wave numbers of the Stokes peak position. However, any strain/stress can also lead to lattice constant changes, thus this modality cannot be used alone in case strain/stress is expected to occur in the structure, in particular upon temperature changes and different coefficients of thermal expansions in a multilayer Au-clad stack. Another approach to determine the temperature is the ratio of the anti-Stokes/Stokes intensities which is not affected by the presence of strain/stress as it is directly linked to the Boltzmann distribution.[38] However, as the intensity of the Stokes' peak relies on the phonon population and if the Raman laser excitation itself creates high photoexcited carriers which induce a large population of non-thermal phonons, the anti-Stokes/Stokes ratio will be directly affected, hence becoming also a less suitable method for our studies. The FWHM of the Stokes peak, however, is also sensitive to temperature as it relies on the phonon distribution: a broadening of the Stokes' peak is expected upon temperature increase.[39]

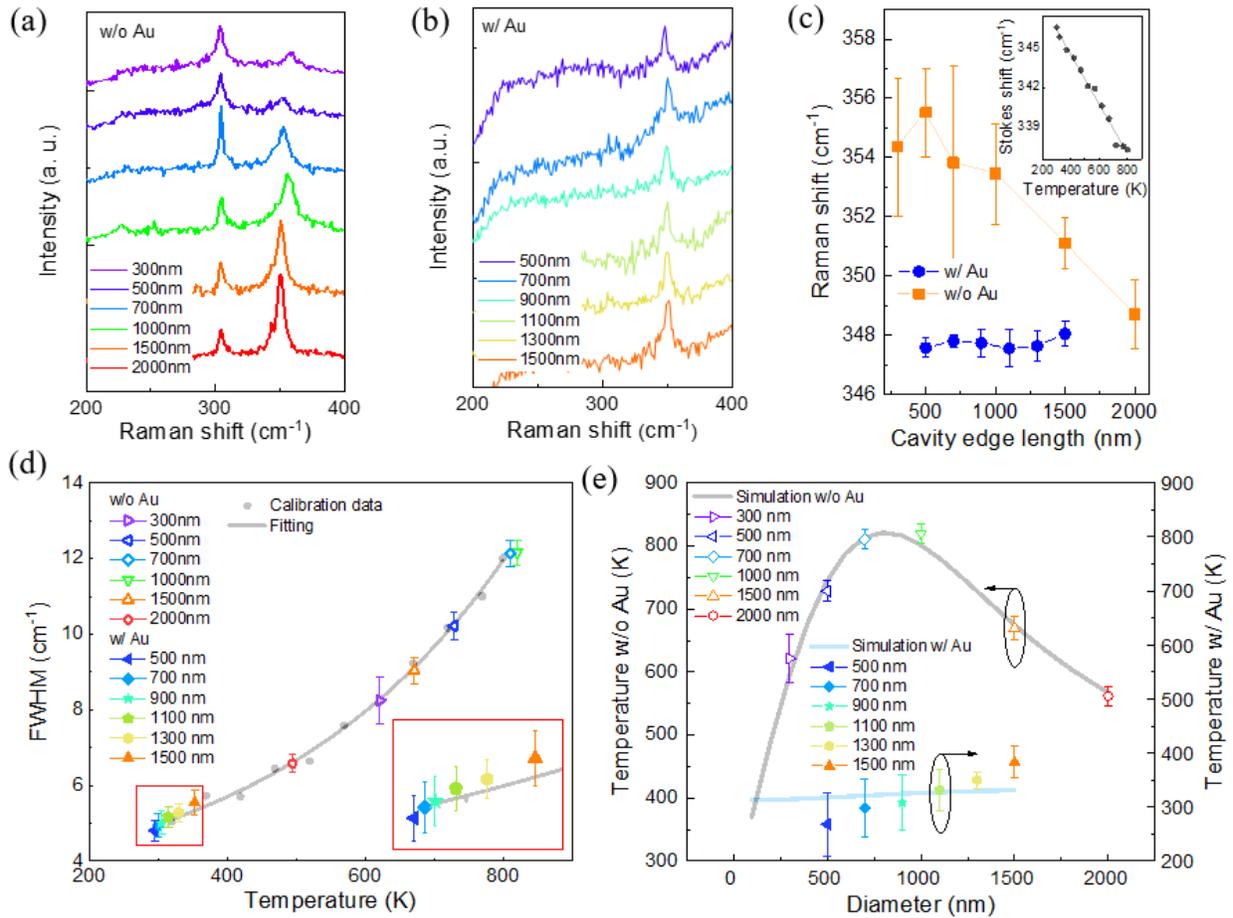

Figure 4. Experimental Raman data on micro- and nanosquare cavities as a function of diameter (square edge length). (a) Raman raw spectra of cavities without Au, obtained with an integration time of 50 s under an excitation laser wavelength of 532 nm at 300 K external temperature. (b) Raman spectra of Au-clad square cavities obtained with an integration time of 200 s under an excitation laser wavelength of 532 nm at 300 K external temperature. (c) Raman Stokes shift of the micro- and nanocavities, inset shows the Raman Stokes shift dependence on temperature acquired on a large rectangle (50 μm × 100 μm) on the same chip acquired by external heating with a Linkam stage over the temperature range from 300 K to 830 K acquired with laser intensity (0.15 mW) below the optical heating threshold. (d) Raman Stokes peak width (FWHM) calibration on the InP rectangle (grey dots and fitting with grey line), obtained within the same temperature range from 300 K to 830 K and with the same laser intensity of 0.15 mW. In addition, the FWHMs of the InP squares are plotted when being excited with 2.31 mW laser power. Inset shows the zoom-in image of the low temperature part. (e) Simulated temperature (solid line) and temperature derived from the peak broadening (points) of micro- and nanocavities with (blue line, pink circles) and without Au (grey line, orange

diamonds). Assuming an absorption coefficient of 0.73 in the InP layer, the Raman results fit well with the simulation results with total heat value of 1.69 mW.

In our InP-on-Si nanocavities, we expect both the presence of strain due to the different materials and nanoscale dimensions as well as a high density of photoexcited carriers in the structure as we are effectively pumping them. Therefore, we expect the Stokes peak broadening to be the most appropriate modality for this temperature study in particular as the Anti-stokes/Stokes ratio approaches 1 for the high temperatures expected and does therefore not provide as accurate temperature extractions as for lower temperature ranges. We use the Raman peak of InP's longitudinal optical (LO) mode, which is reported to be located around 340 $cm^{-1}$ to 350 $cm^{-1}$ for InP bulk,[40] 350 $cm^{-1}$ for InP QWs on Si,[41] and 340 $cm^{-1}$ for InP nanowires.[42]

Fig. 4 (a) shows the Raman spectra of purely photonic nanocavities with various diameters from 300 nm to 2 μm (at 300 K without external heating with the heating stage) acquired when exciting at 532 nm with laser power of 2.31 mW. Two series of peaks are observed: the ones located at around 304 $cm^{-1}$ originate from the Si substrate and are being excited through the InP layer, the other series located at around 350 $cm^{-1}$ results from the InP LO mode. The InP LO Raman signal intensities generally increase as the cavity diameter increases whereas the Si signals follow the opposite trend. Fig. 4 (b) shows the Raman spectra of Au-clad micro- and nanocavities with various diameters from 500 nm to 1.5 μm (at 300 K without external heating with the heating stage) acquired when exciting at 532 nm with laser power of 2.31 mW. In these spectra, only the InP LO peaks are observed while the Si peaks are absent since the Au layer covering the entire surface surrounding the InP squares absorbs the Raman scattered light from the Si buried underneath. A slightly smaller peak width is visible in the Raman spectra for the Au-clad cavities in comparison to the InP bare cavities even at these low laser intensities. To derive the temperature of the micro- and nanocavities upon optical heating (to simulate the optical pumping) from the Stokes peak widths, we first performed a calibration of the Stokes peak shifts and peak widths on a large InP rectangle (50 μm × 100 μm) on the same III-V-on-insulator stack ($SiO_2$ thickness of 2 μm) upon externally heating the entire device from the bottom via the heating stage. For that purpose, the stack was put into a vacuum chamber to reduce convection cooling and was then slowly heated up to 850 K in various steps while recording Raman spectra after a long intermediate waiting time to reach thermal equilibrium, with laser intensities (0.15 mW) below the optical heating threshold. More details on the power dependent Raman data can be found in Supporting Information 5. The Raman Stokes peak shifts of the micro- and nanocavities are shown in Fig. 4 (c) with the inset showing the temperature dependent calibration data on the 50 μm × 100 μm rectangle on the same chip. However, we observed a significant difference between the Stokes peak shift on the cavities from the calibration data. We attribute this difference to the strain present in the micro- and nanocavities. Consequently, we do not use the Raman Stokes peak shift to calibrate the temperature. Instead, we here use the FWHM to do the temperature calibration. The temperature dependent calibration data (grey dots) and fitting curve (grey line) are shown in Fig. 4 (d). The fit shows a quadratic relationship between the peak width and the (externally applied) temperature as expected.[43] This calibration curve is used to derive the temperatures of the InP cavities according to their thermally broadened FWHMs when being optically heated with the Raman laser at 532 nm. The peak widths of the InP on Si micro- and nanocavities plotted in Fig. 4 (d) are determined from several Raman spectra when being optically heated under 2.31 mW (the raw data is plotted in Supporting Information 6). The average FWHMs and their scatter are determined by using Lorentz fitting of the InP LO peaks[44] and plotted with the calibration curve in Fig. 4 (d). The different temperatures of the InP cavities as a function of cavity diameter are plotted together with the simulation data in Fig. 4 (e). Assuming an absorption coefficient of 0.73 in the InP layer (a reasonable value taken both from literature[45] and our optical experiments), both the temperatures from the micro- and nanocavities without Au and the Au-clad cavities agree quantitatively well with the steady-state simulation results. For the purely photonic cavities, the functional behavior follows exactly the one found in the simulations. For the Au-clad cavities, the relatively large variation at the individual cavity size, reflected in the error bar, is due to slight differences of the laser spot position and focusing on the nanocavities. The slightly steeper increase in temperature found in the experiments compared to the simulations may be explained by the fact that the Au claddings might not cover cavities with a perfectly homogeneous film, but will contain grains of various sizes as well as other non-uniformities. In the simulations, however, the Au cladding are ideally covered. These results also comply with the optical performance of the Au-clad nanocavities and the nanocavities without Au in our previous study, where the Au-clad devices show evidence of room temperature lasing with diameter down to 300 nm, while the purely photonic ones show lasing with diameter down to 500 nm under 750 nm optical excitation. Moreover, the purely photonic lasers show photoluminescence saturation at much lower excitation power than the metal-clad lasers, which also implies a better heat dissipation in the metal-clad cavities.[46]

## B. Transient thermal analysis

**Transient simulations.** To investigate the dynamic temperature variation under pulsed condition, we carried out transient thermal simulation on the nanocavity lasers. The thermal boundary condition in the transient simulations is the same as in the steady-state simulation while a peak power of ~840 mW equals to a total average power of 1.69 mW as used for heat generation, this again corresponds to the regime right below lasing threshold. We first studied the influence of the pulse width and cycle time/frequency on the transient temperature in one period.

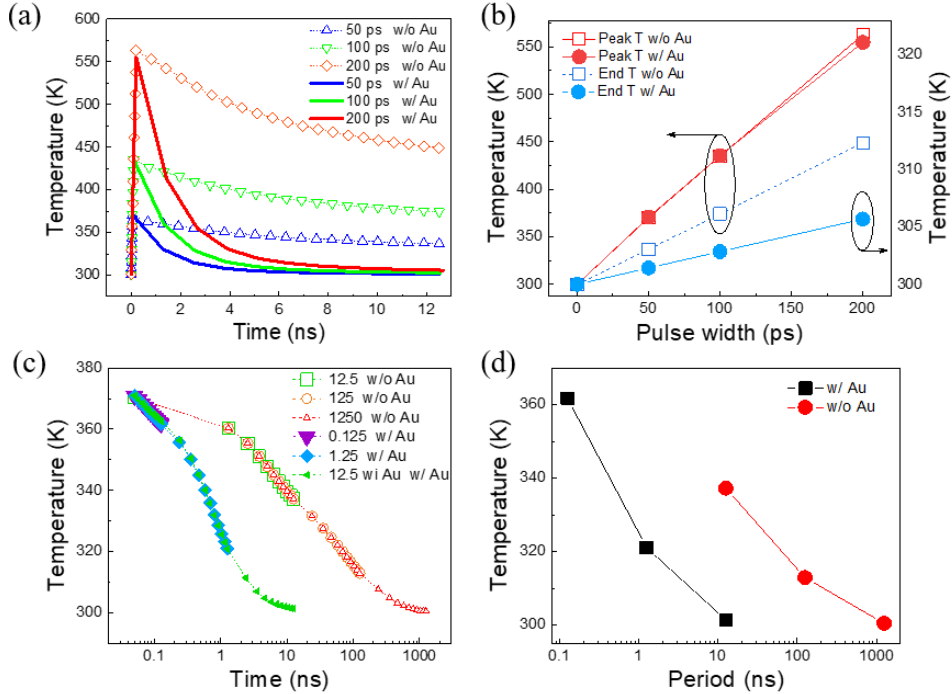

Figure 5. Transient simulation results of a 1 μm disk cavity with a peak power of ~840 mW. (a) Transient temperature of Au-clad and purely photonic cavity in one period with constant cycle time of 12.5 ns and various pulse widths from 50 ps to 200 ps. (b) The peak temperature at the end of the pulse and lowest temperature at the end of the period dependence on pulse width. (c) Transient cooling temperature variation in one period with constant pulse width of 50 ps and various period from 12.5 ns to 1250 ns for cavities without (hollow shapes) Au cladding and from 0.125 ns to 12.5 ns for cavities with Au cladding (solid shapes). (d) Lowest temperature reached at the end of the period for constant pulse width of 50 ps and varying periods for Au-clad (black squares) and purely photonic cavities (red circles).

Fig. 5(a) shows the temperature variation in one period for 1 μm disk cavity with and without Au cladding. Fig. 5(b) shows the peak temperatures at the end of the pulse and the lowest temperatures at the end of the period. For both cavities, the peak temperature and end temperature in one period linearly increases with pulse width. This means that the cavity is illuminated for longer with the same intensity, so in total more energy is provided. We also note from Fig. 5(b) that the highest temperature at the end of the pulse width shows a slightly higher value on the cavity without Au, while the temperature at the end of the period shows a significant higher value of about 145 K for pulse width of 200 ps. We note that the graph contains simulation data extrapolated to faster time scales of 50 ps and below. In this regime, the simulation assumptions do not hold strictly (see Supporting Information 8) but may still serve to estimate general trends. Fig. 5(c) shows the cooling temperature variation for various cycle times (different frequency) while keeping a constant pulse width of 50 ps. The cycle time varies from 12.5 ns to 1250 ns for the cavity without Au while from 0.125 ns to 12.5 ns for the cavity with Au in order to show the entire time range within which the temperature decreases to its initial value. Fig. 5(d) shows the temperature at the end of the period for different cycle times, indicating that a much longer cycle time is required for the cavity without Au to drop to a low temperature close to ambient. This suggests the need for a pulsed operation with much longer cycle time for the microcavity without Au, from a thermal perspective.

Fig. 6(a) shows the transient temperature change of a 1 μm disk cavity in two periods, the inset depicts a schematic pulse injection with cycle time of 12.5 ns and pulse width of 50 ps. A peak power of ~840mW is used in the simulation. Comparing the peak temperature variation for the microdisk cavity in the first period, the highest temperature is almost the same at the end of the pulse, while the lowest temperature at end of one period is about 36 K lower for the cavity with Au cladding. From this, we infer a much smaller time constant of ~10 ns for the nanocavity with Au. This time constant relates to the local heating of the cavity. Taking into account the heating of the surrounding of the cavity, a second, slower time constant emerges. Fig. 6(b) shows the temperature variation over 100 periods. The red curve and the black curve show the transient temperature under pulsed condition while the blue and green curves show the comparative temperature evolution under cw operation, the inset shows the temperature under cw operation in log time scale till reaching the steady-state temperature (with the same total power of 1.69 mW). For the Au-clad cavity, the temperature oscillates between the highest temperature of ~375 K and room temperature, compared to an average temperature increase in the cw case of about 25 K. In the microcavity without Au, the temperature oscillates within a temperature range of 150 K around a medium temperature which is hundreds of Kelvins higher than the room temperature. Both the high temperature spike and large temperature variation under the pulsed operation are considered to be detrimental for the nanocavity without Au.

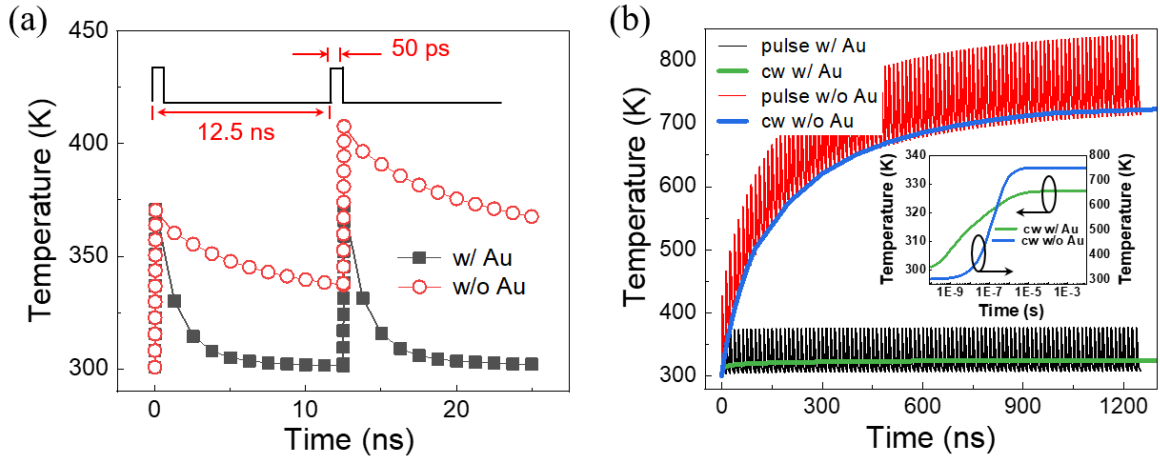

Figure 6. Transient simulation results of a 1 μm disk cavity with a peak power of ~840 mW. (a) Transient temperature change within two periods with pulse width of 50 ps and cycle time of 12.5 ns. (b) Transient temperature change within 100 periods for Au-clad disk cavity and purely photonic disk cavity, inset is the transient temperature with cw power of 1.69 mW for both Au-clad and purely photonic cavities, green curve for Au-clad cavity and blue curve for purely photonic cavity.

**Experimental photoluminescence results under pulsed operation.** In order to evaluate the transient simulation results on microcavity lasers, we measured the photoluminescence (PL) of 2 μm square cavities with ten 5.5 nm thick strain-compensated $In_{0.72}Al_{0.11}Ga_{0.17}As$ QW and $In_{0.44}Al_{0.32}Ga_{0.24}As$ barriers embedded in InP with a pulsed laser centered at 1064 nm. The structure is created by wafer-bonding of the 300 nm thick III-V layer on top of a Si wafer coated with 2 mm thick $SiO_2$. In this experiment we evaluate only microcavities without Au, as this is where we expect to see the largest effects in our simulation results. According to the simulations, the thermal effects of the InP cavities with InAlGaAs quantum wells are transferrable to InP bulk cavities. The results are shown in Supporting Information 7. The reason for using a QW structure for this experiment, as opposed to pure InP cavities, is the 1064 nm emission wavelength of our cw pump source, which would not be sufficiently absorbed in the previously discussed InP structures. This pump laser is capable of both cw and pulsed operation with the capability of changing working current, frequency and pulse width.

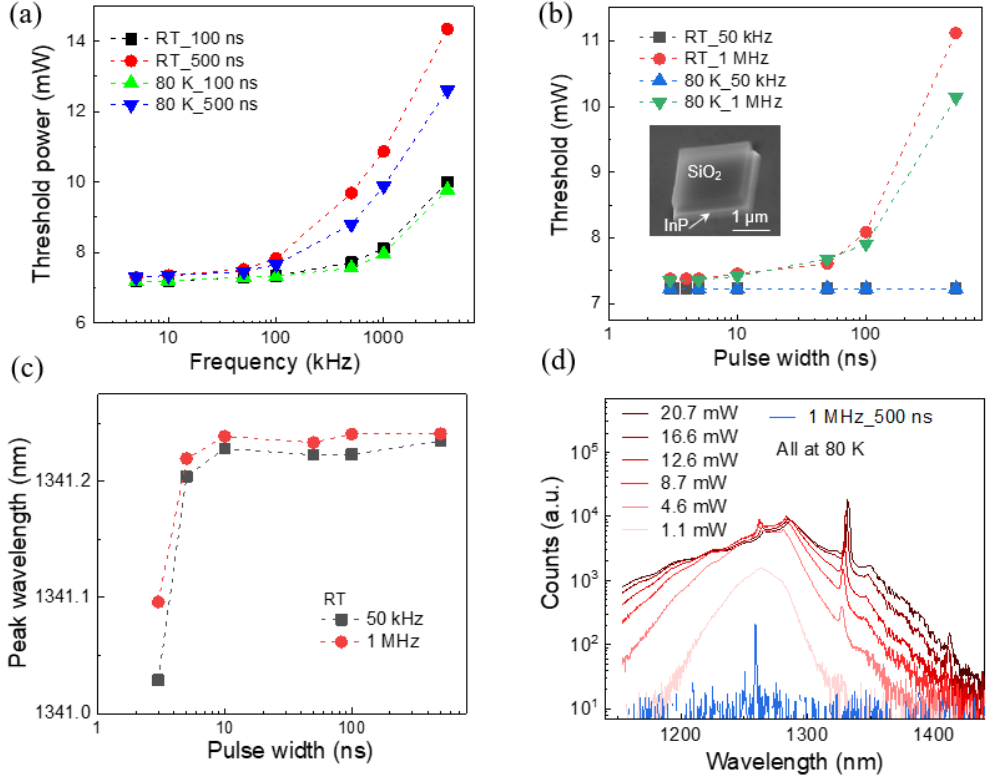

Figure 7. Experimental optical results. (a) Frequency-dependent threshold power of a 2 μm purely photonic square laser under various temperatures and pulse widths. The black squares and red dots are at room temperature (RT) with pulse width of 100 ns and 500 ns, while the green and blue triangles are at 80 K with pulse with of 100 ns and 500 ns, respectively. (b) Pulse width-dependent threshold power of the 2 μm purely photonic square laser under RT with frequency of 50 kHz and 1 MHz (black square and red dots) and at 80 K with frequency of 50 kHz and 1 MHz (blue and dark green triangles). Inset shows the SEM image of the micro-square laser. (c) Peak wavelength dependence on pulse width showing a red-shift of the PL peak as pulse width increases. (d) PL spectra of the 2 μm purely photonic square laser at 80 K with various cw laser power (red) and lasing peak at 80 K under pulsed pumping with frequency of 1 MHz and pulse width of 500 ns.

Fig. 7 (a) shows the threshold power dependence on frequency with fixed pulse width of 100 ns and 500 ns under room temperature (RT) and 80 K. As the frequency increases, the threshold power of the micro-square laser shows a superlinear increase. Fig. 7 (b) shows the threshold power dependence on pulse width, where a similar superlinear increase is observed as the pulse width increases. Inset is the SEM image of the measured micro-square laser with a diameter of 2 μm. The increase of the threshold optical power with either increase of frequency or pulse width can be attributed to the temperature increase with the increase of the frequency or pulse width. Fig. 7 (c) shows the PL peak wavelength at room temperature under frequency of 5 kHz and 1 MHz. Red-shifts of the PL peak are observed under both frequencies with pulse width increases. Fig. 7 (d) shows the spectra of the 2 μm square cavity at temperature of 80 K with different cw optical power from 1.1 mW to 20.7 mW (in red) together with the spectra under pulsed condition with frequency of 1 MHz and pulse width of 500 ns (in blue). Taking both the experimental results and the simulation results into consideration, we conclude that a pulsed pumping with frequency of 100 kHz and pulse width of 100 ns is the optimal operation mode for the 2 μm purely photonic micro-square cavity from thermal perspective.

DISCUSSION

In conclusion, we studied the heating effects of III-V-on-Si micro- and nanocavity lasers under optical pumping. We investigated the impact of a Au cladding on device temperature. Our simulation results show a drastic temperature decrease of several hundreds of Kelvins in nanocavity lasers with Au cladding compared to purely photonic cavities. We validated the heating effects of nanocavities experimentally using Raman thermometry. The Raman Stokes peak width broadening results agree well with simulations. Moreover, by comparing the temperature with different cavity diameters and shapes, the highest temperatures are observed on the purely photonic cavities with diameter around 700 nm, as a result of the Gaussian profile of a 1 μm pump laser. The thickness of the bottom $SiO_2$ and surrounding $Al_2O_3$ were also studied, showing that the thickness of the underlying $SiO_2$ plays a significant role, and decreasing the thickness can help to reduce the temperature. However, a lower boundary of the SiO$_2$ layer thickness is given by the dielectric confinement of the mode, but it should be possible to reduce this to about 300 nm, based on previously demonstrated results.[24] The thickness of Al$_2$O$_3$, on the other hand, does not have much influence on the device temperature profile.

In this study, we also evaluated the heating effects of the pulsing scheme both by simulation and experimentally. The transient simulation results confirm the fast and efficient heat dissipation in the Au-clad cavity. While the temperature of both the purely photonic and the Au-clad cavity increases as the pulse width and frequency increase, the resulting temperature increase is much higher for the purely photonic cavities. To compare this with experimental data, we measured the threshold power of In-AlGaAs QW microcavity lasers in both cw and pulsed mode. A pulsed condition with frequency of 100 kHz and pulse width of 100 ns is the optimal operation mode for the 2 μm square laser from a thermal perspective.

This study provides guidelines for nanocavity design and operation based on thorough thermal simulations and experimental study by using of Raman characterization and photoluminescence. We believe that our findings improve the understanding of thermal effects in nanocavity lasers. Although we investigate here specifically InP-on-Si nanocavities, we believe our methods could be transferable to other photonic devices and platforms to help guide others in device design. In particular, while the use of metal cavities has been proposed mainly to enable downscaling of nanocavity lasers beyond the diffraction limit by the use of hybrid photonic-plasmonic modes, here we show that - beyond such consideration - the use of metals plays a significant role in reducing the operating temperature of nanocavity lasers and that this positive effect might off-set the inevitable increase in optical threshold power induced by the presence of metal in close proximity to the lasing mode.

METHODS

**Thermal simulation.** Thermal simulations are carried out using commercial finite element method software (ANSYS Parametric Design Language). In the simulation, initial temperature of the device is set to be uniform of 300 K. The backside of Si substrate is set to be constant of 300 K. Detailed considerations on heat convection in the simulation is discussed in the Supporting Information 1. The heat transfer of the nanocavity lasers is modeled uses Fourier's law:[47]

$$q = -\kappa \cdot \nabla T \quad (1)$$

where q is the heat flux, κ is the thermal conductivity and ∇T is the thermal gradient between the heat source and the heatsink. The justification for a thermal transport modelling based on diffusive transport is detailed in the Supporting Information 8. The mean free path of InP is calculated to be 65 nm, which is smaller than all linear dimensions. According to experimental results in literature, we could neglect the thermal conductivity dependence on the phonon mean free path in SiO$_2$ (with phonon mean free path less than 60 nm[48]). Although the mean free path of charge carriers in Au is relatively large (~38 nm[49]), it is still much smaller than the smallest dimension used in the simulation. Though the phonon mean free path in Si substrate is about 300 nm at 300 K,[50] the dimensions of the Si substrate used in the simulation is 50 μm × 50 μm × 50 μm. Therefore, we use the typical thermal conductivity values shown in Table 1.

The transient simulation is based on the following three-dimensional heat conduction equation:[51, 52]

$$\frac{\partial T}{\partial t} = \kappa \left\{ \frac{\partial^2 T}{\partial x^2} + \frac{\partial^2 T}{\partial y^2} + \frac{\partial^2 T}{\partial z^2} \right\} + \frac{Q}{\rho c} \quad (2)$$

Where T is the temperature, κ is the thermal conductivity, ρ is the density, c is the specific heat of the material, and Q is the heat generation density.

Optical pumping is modelled through a spatially-dependent heat generation density Q (x, y) in the InP layer. We adopt a gaussian distributed value with a total integration power of 1.690 mW. This corresponds to the situation of slightly below the lasing threshold for the majority of our devices, so we assume the total optical power is converted into heat. As the absorption is assumed to be constant along the thickness direction of the InP layer, here we use a gaussian distributed heat generation rate $H_{gen}$ with the integration value equals to $H_{gen}$=Q/t$_{InP}$, where t$_{InP}$ is the thickness of the InP layer. Taking the spot size of the pumping laser (1 μm) into account, we applied the heat generation rate, $H_{gen}$, as follows:

$$H_{gen} = A * \exp(-B * (x^2 + y^2)) \quad (3)$$

where A is 1.990e16 W/m$^2$, B is 1.109e13. The unit of the $H_{gen}$ is W/m$^2$.

**Raman thermometry characterization.** The Raman measurements are performed under ambient in the IBM Noisefree lab with a confocal LabRam HR tool from Horiba Jobin Yvon. The Raman spectra are acquired with a 100 × objective with 0.86 numerical aperture and a laser source with 532 nm excitation wavelength. The laser spot size is < 1 μm which gives a high spatial resolution. The temperature dependent calibration is carried out in a Linkam vacuum chamber (approx. 5 × 10$^{-1}$ mbar) with the capability of varying temperature from 78 K to 900 K thanks to liquid nitrogen cooling. Fixed laser power of 0.15 mW is used and integration time of 50 s and 200 s are used for the purely photonic cavities and Au-clad cavities, respectively. During the acquiring of the Raman spectra, a 1800 gr/mm grating was used which gives high-resolution grating with 0.3 cm$^{-1}$.

**Optical threshold characterization.** The optical photoluminescence (PL) measurements are performed under cw and pulsed operation at 80 K and 300 K. The device is placed in a cryostat where light is focused on the device using an objective with a magnification of 100 × and a numerical aperture of 0.6 placed inside the cryostat. The PL emission is collected by the same objective and detected by an InGaAs line array detector which is combined with a grid diffraction spectrometer. The pump laser is centered around 1064 nm with capability of operating under cw or pulsed mode. The pulse width can be tuned from 100 ps to 500 ns and frequency can be tuned from 1 Hz to 4 MHz. In this study we only show PL measurements of InP/InAlGaAs QW devices because of the fixed wavelength of our cw pump laser, which is not suitable for the pumping of InP cavities. In earlier work we evaluated the lasing behavior of metal-clad and purely photonic devices and we refer to those works for more details.[42] The excitation laser is operated by varying the current. To map the average excitation power to driving currents of the source, the power is measured with a power meter.

ASSOCIATED CONTENT

**Supporting Information**
Steady-state simulation results showing temperature distribution on each part; Steady-state results with various Au cladding dimensions; Steady-state results with various Al$_2$O$_3$ thickness; Comparison of steady-state simulation results with constant thermal conductivity and temperature dependent thermal conductivity; Power dependent Stokes peak shift and Stokes peak width; Raw data of experimental Stokes peak shift and Stokes peak width; Comparison of thermal effects in InP bulk cavities and quantum well cavities; Applicability of simulation results in terms of length scale and time scale.

AUTHOR INFORMATION

**Corresponding Author**
Pengyan Wen - IBM Research Europe - Zurich, Saeumerstrasse 4, 8803 Rueschlikon, Switzerland; Email: pew@zurich.ibm.com.

**Authors**
Preksha Tiwari - IBM Research Europe - Zurich, Saeumerstrasse 4, 8803 Rueschlikon, Switzerland. She is now a postdoc at Paul-Scherrer Institute.

Markus Scherrer - IBM Research Europe - Zurich, Saeumerstrasse 4, 8803 Rueschlikon, Switzerland.
Emanuel Lörtscher - IBM Research Europe - Zurich, Saeumerstrasse 4, 8803 Rueschlikon, Switzerland.
Bernd Gotsmann - IBM Research Europe - Zurich, Saeumerstrasse 4, 8803 Rueschlikon, Switzerland.
Kirsten E. Moselund - IBM Research Europe - Zurich, Saeumerstrasse 4, 8803 Rueschlikon, Switzerland. She has since moved to a joint position with the Paul-Scherrer Institute and EPFL, but this work was carried out entirely at IBM.

**Author Contributions**
P. T. fabricated the devices. P. W. did the thermal simulation. P. W. and B. G. performed the thermal analysis. P. W. and E. L. conducted Raman spectroscopy and analysis. P. W. and P. T. and M. S. performed the optical characterization. P. W. wrote the article with contributions from all authors. K. M. led the project.

**Acknowledgements**
We gratefully acknowledge technical support for material growth from Heinz Schmid, for device fabrication from Svenja Mauthe, Yannick Baumgartner and for CMP from Daniele Caimi and we thank the Cleanroom Operations Team of the Binnig and Rohrer Nanotechnology Center (BRNC) for their help and support.

**Funding Sources**
This work has received funding from the European Union H2020 ERC Starting Grant project PLASMIC (Grant Agreement No. 678567) and H2020 MSCA IF project DATENE (Grant Agreement No. 844541).

**Notes**
The authors declare no competing financial interests.